# Distributed cladding mode fiber-optic sensor

Gil Bashan, Yosef London, H. Hagai Diamandi, and Avi Zadok*

*Faculty of Engineering and Institute for Nano-Technology and Advanced Materials, Bar-Ilan University, Ramat-Gan 5290002, Israel*
*Corresponding author: Avinoam.Zadok@biu.ac.il*

The analysis of substances outside the cladding has challenged the optical fiber sensors community for decades. A common solution relies on the propagation of light in cladding modes. However, the coupling of light to/from these modes is typically based on permanent gratings in specific locations, which restrict the sensors to point measurements only. In this work, we present dynamic, random-access coupling of light between core and cladding modes of standard fibers, in arbitrarily located short sections. Coupling is based on the stimulation of Brillouin dynamic gratings by two coded pump waves and intermodal scattering of a third optical probe wave. All waves are launched and collected in the core mode. No permanent gratings are inscribed. Distributed sensing of surrounding media is demonstrated with 2 m range and 8 cm resolution. Measurements distinguish between water and ethanol outside the fiber. The measurement accuracy of the local index outside the cladding is 0.004–0.0004 refractive index units.　© 2020 Optical Society of America under the terms of the OSA Open Access Publishing Agreement

https://doi.org/10.1364/OPTICA.377610

## 1. INTRODUCTION

Optical fibers constitute an exceptional sensing platform [1–3]. They reach measurement ranges of hundreds of kilometers, are simply embedded inside many structures, are comparatively immune to electromagnetic interference, and are suitable for installation in harsh environments [1,2]. Furthermore, fibers support spatially distributed mapping of quantities of interest, in which every segment serves as an independent sensor node [3–9]. However, the analysis of media outside the fiber boundary has represented a fundamental challenge to the optical fiber sensors community for decades. Standard single-mode fibers are designed to guide light within an inner core and minimize any leakage, whereas a substance under test typically lies outside the boundary of the fiber cladding and even outside the protective coating.

All standard optical sensing protocols, such as refraction, absorption, scattering, or fluorescence mandate spatial overlap between light and test media. Therefore, most optical fiber sensors of chemicals and biological reagents rely on nonstandard geometries such as hollow-core photonic crystal fibers [10–13], specialty reactive materials [14–17], or considerable structural modifications of the standard fibers [18–24]. Over the last three years, a breakthrough in distributed analysis of liquids outside the fiber has been achieved based on fiber optomechanics: the coupling between light and guided acoustic modes of the entire cladding cross section [25–32]. Recent demonstrations of distributed optomechanical fiber sensors reached a spatial resolution between 15–50 m [28,29], limited by long acoustic lifetimes and/or poor signal-to-noise ratios.

One of the most widely employed protocols for fiber-optic chemical sensing relies on the propagation of light in cladding modes [33–38]. The transverse profiles of these modes are characterized by evanescent tails, which overlap with media outside the cladding. Sensors based on cladding modes have been reported in hundreds of works over more than 20 years [33–38]. The main challenge in such sensors is the coupling of light to/from the cladding modes. The most common solution path is based on fiber gratings: A long-period grating can couple light between the core mode and a copropagating cladding mode [33,38], whereas a short-period fiber Bragg grating may do the same for a counterpropagating cladding mode [34,38]. The coupling efficiency may be enhanced using gratings that are tilted at an angle [39–41]. However, both schemes suffer from a fundamental drawback: they require the inscription of permanent perturbations in the fiber. Consequently, cladding-mode fiber sensors have been almost exclusively restricted to point measurements, and their extension towards spatially distributed analysis has remained difficult.

Over the last decade, a significant alternative to the inscription of permanent fiber gratings has been established in the form of Brillouin dynamic gratings (BDGs) [42–52]. A pair of counter-propagating optical pump fields in one spatial mode of the fiber is used to generate an acoustic wave through a backwards-stimulated Brillouin scattering (SBS) process [42,52]. The acoustic wave is accompanied by a traveling grating of refractive index perturbations, due to photoelasticity [42,52]. Unlike standard gratings, which are based on local structural modifications [53], BDGs may be switched on and off at will. Moreover, careful modulation of the two pump waves may confine the stimulation of BDGs to discrete and narrow fiber sections and scan their positions along the fiber [44–46,52].

Once established, BDGs may also affect additional optical probe fields, which propagate in different spatial modes in the same fiber. The concept is most often implemented in polarization-maintaining fibers: Pump waves polarized along one principal axis generate the BDG, and a probe wave of the orthogonal polarization is reflected by the grating [42]. BDGs over polarization-maintaining fibers are used in distributed sensing of strain and temperature, variable all-optical delay lines, microwave photonic filters and all-optical signal processing applications [42–52]. The principle is not restricted to polarization-maintaining fibers only: in one notable example, Kim and Song used pump and probe waves in different guided core modes of a few-order-mode fiber [54]. However, BDGs were not yet employed in the coupling of light to cladding modes.

In this work, we report the dynamic, random-access coupling of light between counterpropagating core and cladding modes in a standard single-mode fiber. Coupling is based on BDGs that are confined to few-centimeter long sections arbitrarily located along the fiber. All optical fields are launched and collected in the core mode only. No permanent gratings are necessary. The principle is used in a first demonstration of a distributed cladding-mode fiber-optic sensor. The refractive index of media outside the cladding is mapped over 2 m of bare fiber, with spatial resolution of 8 cm. Changes in local coupling spectra properly identify short sections of fiber immersed in water and ethanol and clearly distinguish between the two liquids. The local refractive index outside the cladding is estimated with an accuracy between 4e-3 to 4e-4 refractive index units (RIUs), depending on base value. The measurements establish a new class of distributed fiber sensors. Results were briefly reported in a recent conference presentation [55].

## 2. RESULTS

### A. Principle of Operation

Consider two copolarized optical pump waves $E_{\text{pump}1,2}$ that counterpropagate in the core mode of a standard fiber. The central optical frequencies of the two waves are $\omega_{\text{pump}1}$ and $\omega_{\text{pump}2} = \omega_{\text{pump}1} - \Omega$, respectively, where the offset $\Omega$ is close to the Brillouin frequency shift $\Omega_B$ of the fiber. We denote the effective index of the single core mode at all frequencies of interest as $n_{\text{core}}$. The two pump waves generate an acoustic wave of frequency $\Omega$ via SBS, which copropagates with $E_{\text{pump}1}$. The dispersion relations of the optical core mode in both directions and that of the acoustic mode are illustrated in Fig. 1(a).

Consider next an optical probe wave $E_{\text{probe}}$ of optical frequency $\omega_{\text{probe}} > \omega_{\text{pump}1}$ that is copropagating with $E_{\text{pump}1}$ in the core mode [Fig. 1(b)]. As shown in Figs. 1(a) and 1(b), the BDG induced by the two pumps can couple between the probe wave and a fourth optical field $E_{\text{clad}}^{(m)}$ of frequency $\omega_{\text{clad}} = \omega_{\text{probe}} - \Omega$, which counterpropagates in the $m$th-order cladding mode of the fiber. We denote the effective index of the clad mode as $n_{\text{clad}}^{(m)} < n_{\text{core}}$. Calculations of $n_{\text{clad}}^{(m)}$ as a function of mode order in a 125-μm-diameter fiber are shown in Supplement 1, Fig. 1. Phase-matched coupling to cladding mode $m$ is achieved for a specific optical frequency of the probe wave (see Supplement 1 for a detailed derivation),

$$\omega_{\text{probe,opt}}^{(m)} = \frac{2n_{\text{core}}\omega_{\text{pump}1} - \left(n_{\text{core}} - n_{\text{clad}}^{(m)}\right)\Omega}{n_{\text{core}} + n_{\text{clad}}^{(m)}}$$

$$\approx \frac{2n_{\text{core}}}{n_{\text{core}} + n_{\text{clad}}^{(m)}} \omega_{\text{pump}1}. \quad (1)$$

The frequency difference $\Delta\omega$ between the copropagating pump and probe at optimal phase matching is given by

$$\Delta\omega_{\text{opt}}^{(m)} \equiv \omega_{\text{probe,opt}}^{(m)} - \omega_{\text{pump}1} \approx \frac{n_{\text{core}} - n_{\text{clad}}^{(m)}}{n_{\text{core}} + n_{\text{clad}}^{(m)}} \omega_{\text{pump}1}. \quad (2)$$

Calculated wavelength offsets corresponding to $\Delta\omega_{\text{opt}}^{(m)}$, for a pump wave at 1550 nm, are shown in Supplement 1, Fig. 1.

The coupling of power between the core and cladding modes is very weak (see also below). At that limit, the BDG-induced optical power reflectivity may be approximated by (see Supplement 1, [56,57]),

$$R^{(m)}(\Omega, \Delta\omega) = \frac{D_{\text{BDG},0}^2 |Q_{\text{core}}|^2 \left|Q_{\text{clad/core}}^{(m)}\right|^2 P_{\text{pump}1} P_{\text{pump}2} L_{\text{BDG}}^2}{1 + [2(\Omega - \Omega_B)/\Gamma_B]^2}$$

$$\times \text{sinc}^2\left(\frac{\Delta k_{\text{BDG}}^{(m)} L_{\text{BDG}}}{2}\right). \quad (3)$$

In Eq. (3), $L_{\text{BDG}}$ denotes the length of the BDG, $P_{\text{pump}1,2}$ (W) are the input optical power levels of the two pump waves, which are assumed to be undepleted along the fiber, and $D_{BDG,0}$ (in m × W$^{-1}$) is a coupling coefficient that depends on optical, acoustic, and photoelastic parameters of silica [42,52]. Also in the same equation, $Q_{core}$ (m$^{-1}$) is an overlap integral between the transverse profiles of the optical core mode and an acoustic mode that is guided in the core of the fiber, and $Q_{clad/core}^{(m)}$ (m$^{-1}$) is an overlap integral between the transverse profiles of the core optical mode, the $m$th-order cladding mode, and the acoustic mode (see Supplement 1). Lastly, $\Gamma_B \approx 2\pi \times 30$ MHz represents the Brillouin linewidth in silica and $\Delta k_{BDG}^{(m)} \equiv (n_{core} + n_{clad}^{(m)})(\Delta\omega - \Delta\omega_{opt}^{(m)})/c$ is a wavenumber mismatch term associated with the probe wave frequency detuning. Here $c$ stands for the speed of light in vacuum. The BDG coupling to the cladding mode is maximal when the frequency detuning between the pump waves matches the Brillouin shift exactly ($\Omega = \Omega_B$), and the probe frequency is tuned to its optimum value above ($\Delta\omega = \Delta\omega_{opt}^{(m)}$). The linewidth of BDG coupling with respect to $\Omega$ follows $\Gamma_B$. The linewidth with respect to $\Delta\omega$ is inversely proportional to the BDG length. For 50-cm-long uniform gratings, for example, the full width at half-maximum is $2\pi \times 190$ MHz.

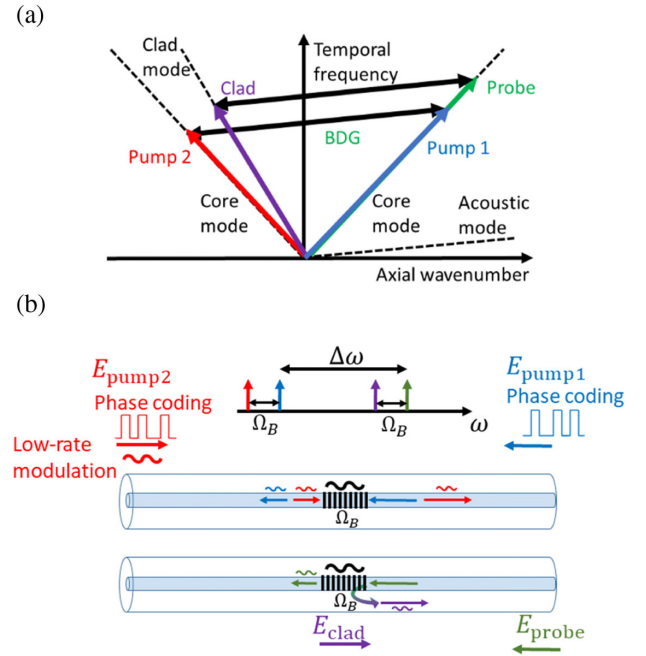

**Fig. 1.** Coupling between counterpropagating core and cladding modes through stimulated BDGs. (a) Phase matching considerations. Illustration of the dispersion relations of the core optical mode of a single-mode fiber in both directions, a cladding mode in one direction, and an acoustic mode excited by backward SBS between two counterpropagating pump waves in the core mode. BDG coupling between a probe optical field in the core mode and a counterpropagating field in the cladding mode is illustrated. (b) Coupling to a cladding mode. Top panel, two phase-coded, counterpropagating pump waves in the core mode (blue and red) stimulate a local BDG (black). The frequency difference between the pump waves is set to the Brillouin shift of the fiber $\Omega_B$. The intensity of one pump is modulated by a low-frequency sine wave (see also Fig. 3). The dynamic grating is therefore modulated at the same rate. Bottom panel, a continuous probe wave in the core mode (green) is launched with specific frequency offset $\Delta\omega$ with respect to the pumps. The probe is scattered by the dynamic grating into a counterpropagating cladding mode (purple). Coupling is accompanied by intensity modulation of the output probe wave, following that of the grating.

Unlike the core mode, the effective indices of cladding modes $n_{clad}^{(m)}$ change with the refractive index $n_{ext}$ of the substance outside the cladding. Therefore, measurements of $\Delta\omega_{opt}^{(m)}$ can be used for sensing outside the fiber. Figure 2(a) presents the anticipated changes in $\Delta\omega_{opt}^{(17)}$ as a function of $n_{ext}$. For example, $\Delta\omega_{opt}^{(17)}$ is reduced by $2\pi \times 2.1$ GHz when a bare fiber is immersed in water ($n_{ext} = 1.316$ RIU at 1550 nm wavelength). The spectral offset of the coupling peak reaches $2\pi \times 3.0$ GHz if the fiber is placed in ethanol instead ($n_{ext} = 1.352$ RIU). Figure 2(a) suggests that the sensitivity of $\Delta\omega_{opt}^{(17)}$ to changes in $n_{ext}$ increases when the index of the surrounding medium approaches that of silica. Note that unlike standard Brillouin sensors, the Brillouin frequency shift $\Omega_B$ remains unchanged, since $n_{core}$ is unaffected by the media under test.

The strength of coupling depends on the mode order $m$ through the transverse overlap between the core, cladding, and acoustic modes. Compared with BDGs in polarization-maintaining fibers, the maximum reflectivity $R^{(m)}(\Omega_B, \Delta\omega_{opt}^{(m)})$ is reduced by a transverse efficiency factor $\eta^{(m)} \equiv |Q_{clad/core}^{(m)}|^2/|Q_{core}|^2$. Figure 2(b) presents numerical calculations of $\eta^{(m)}$ as a function of cladding mode order (see also Supplement 1, Fig. 2). The analysis suggests that coupling is most efficient for odd cladding mode orders between 13 and 19 [34]. Even for these modes, however, $\eta^{(m)}$ only reaches 1.5%. The power reflectivity of a few centimeters-long BDGs written by pump power levels of a few watts is expected to be on the order of only 100 ppm. Nevertheless, the coupling spectra may still be measured and used towards sensing, as discussed below.

Continuous pump waves stimulate BDGs over the entire lengths of fibers under test. Distributed sensing, however, requires that the Brillouin interactions be confined to discrete and narrow fiber sections of arbitrary locations. Several schemes for the localization of steady-state SBS have been proposed towards distributed sensing of temperature and strain [58–62]. In those protocols, which came to be known as Brillouin optical correlation domain analysis (B-OCDA), the two input optical waves are jointly modulated in frequency or phase [58–62]. The concept was successfully carried over to BDGs in polarization-maintaining fibers [44–46,52]. In this work, we modulate the phases of both pump waves by a carefully constructed binary sequence with a period of $N$ bits and short symbol duration $T$ [46,52,60,61]. Modulation effectively confines the BDGs to discrete correlation peaks of width $L_{BDG} = \frac{1}{2}v_g T$, where $v_g$ is the group velocity of light in the fiber [46,52,60,61].

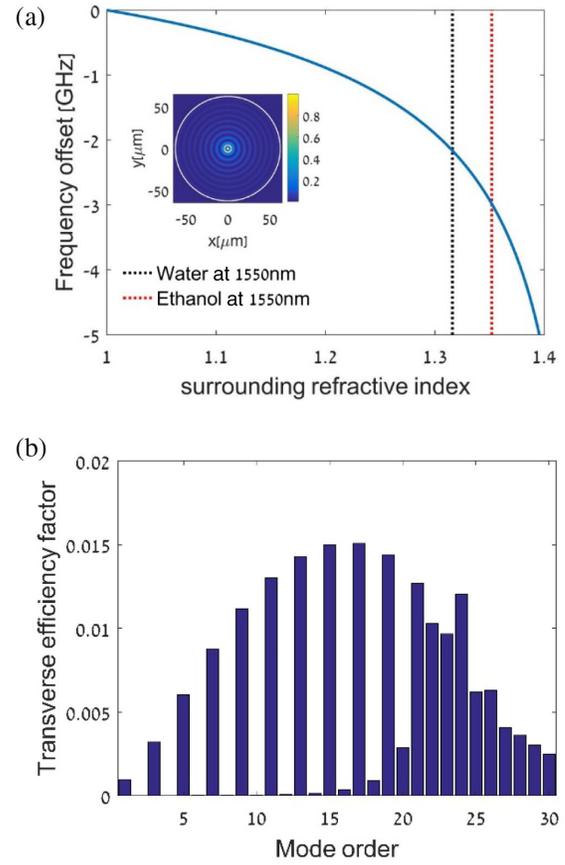

**Fig. 2.** (a) Calculated shift in the probe frequency of maximum coupling to the 17th-order cladding mode as a function of the refractive index of the medium outside the cladding. The refractive indices of water and ethanol at 1550 nm wavelength are noted. The inset shows the transverse profile of intensity in the cladding mode. (b) Calculated transverse efficiency factor of BDG coupling as a function of cladding mode order. The efficiency of BDG coupling between counterpropagating optical probe waves in the core mode, such as in polarization maintaining fibers, is denoted as unity. Coupling to the cladding modes is the most efficient for odd mode orders between 13 and 19. Even for those modes, however, the relative efficiency is 0.015 at most.

The magnitude of the BDG reaches its steady-state value within the correlation peak and oscillates with zero mean everywhere else [61]. Therefore, time-averaged measurements of coupling spectra may be related to the conditions prevailing at a specific position. Location becomes ambiguous when the length of fiber under test exceeds $NL_{BDG}$. However, the sequence length $N$ can be arbitrarily long. Careful retiming of the modulation of the two pump waves may scan the position of the correlation peak over the length of the fiber [61]. Therefore, BDGs induced by phase-coded pumps would provide dynamic, random-access coupling between core and cladding modes in arbitrarily located, short fiber segments. Phase coding in SBS is addressed with more detail in Supplement 1 and is discussed at length in numerous references [46,52,60–62]. The technique serves in distributed sensing of media outside the fiber cladding with a few-centimeter spatial resolution, as described next.

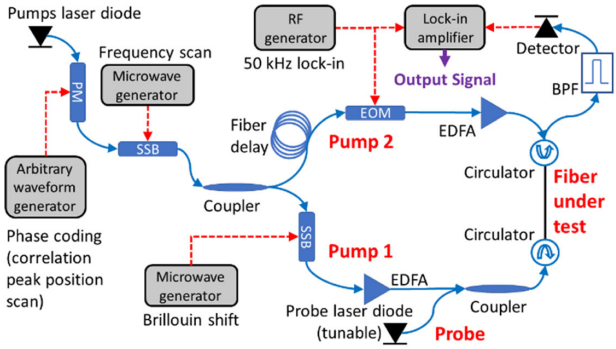

**Fig. 3.** Experimental setup. Schematic illustration of the measurement setup used in distributed sensing of media outside the cladding of a standard bare fiber, based on BDG coupling to cladding modes [55]. SSB, single-sideband modulator; EOM, amplitude modulator; PM, phase modulator; BPF, optical bandpass filter; EDFA, erbium-doped fiber amplifier; RF, radio frequency. Fiber paths are noted in solid blue lines, and electrical cables are marked in dashed red lines.

### B. Experimental Setup and Results

The experimental setup used in distributed fiber sensing based on dynamic coupling to cladding modes is illustrated in Fig. 3. Light from a laser diode of 1554.77 nm wavelength was used as the source of both pump waves. The laser output passed through an electro-optic phase modulator, driven by the output voltage of an arbitrary waveform generator. The generator was programmed to repeatedly apply a 43-bit-long binary sequence, with a symbol duration of 800 ps. Phase coding restricted the Brillouin interaction between the two pumps to correlation peaks of 8 cm width [61,62]. The optical frequency of the pumps source could be scanned in a few megahertz increments using a suppressed-carrier single-sideband (SC-SSB) electro-optic modulator, driven by the output of a microwave generator. The spectral shifts provided fine-tuning of the difference $\Delta\omega$ between the optical frequencies of the pump waves and that of the probe field (see below).

Pump light was split into two paths. The pump waveform in one branch was upshifted in frequency using a second SC-SSB modulator, driven by a sine wave from the output of a second microwave generator. The offset frequency $\Omega$ was chosen near the Brillouin frequency shift of the fiber under test: $\Omega_B \approx 2\pi \times 10.715$ GHz. The offset waveform was then amplified by an erbium-doped fiber amplifier to an average optical power of 35 dBm, and launched into one end of a bare fiber under test as $E_{pump1}$. The waveform in the other branch was amplitude-modulated by a sine wave of 50 kHz frequency. The strengths of the stimulated BDGs were therefore modulated at the same rate. The waveform was delayed over 700 m of fiber, to allow for precise scanning of the BDG position (see below, [63]). Last, the optical wave was amplified in a second fiber amplifier to an average power of 30 dBm and launched into the opposite end of the fiber under test as $E_{pump2}$.

The position of the correlation peak was scanned over the length of the fiber under test as follows [63]. The fiber paths of the two pump waves connect through the short fiber under test to form a loop (see Fig. 3). Let us denote the middle position of that fiber loop as $z_0$. Phase modulation of the two pumps gives rise to BDGs at discrete correlation peaks, located at positions $z_M = z_0 + MN \cdot \frac{1}{2}v_g T$, where $M$ is a positive or negative integer. Due to the large imbalance between the fiber paths of the two pumps, the single correlation peak which overlaps with the short test fiber of interest corresponds to a high order: $M \gg 1$. The exact location of that peak is conveniently adjusted through small-scale changes in the coding symbol duration $T$ [63].

Continuous probe light at optical frequency $\omega_{probe}$ was drawn from a second laser diode of $-4$ dBm optical power. The wavelength of the laser diode could be tuned in 8 pm increments (1 GHz frequency steps), over several nanometers. The probe was launched into the fiber under test in the same direction as $E_{pump1}$. The BDG induced coupling of probe light into counterpropagating cladding modes, as discussed above. Unlike BDGs over polarization-maintaining fibers, the direct observation of reflected probe light in the cladding modes is difficult. Transmission losses of the probe wave in the core mode were therefore monitored instead. Coupling with the BDG resulted in a weak 50 kHz modulation of the transmitted probe optical power. The probe wave was detected at the opposite end of the fiber under test, and the photocurrent was monitored by a lock-in amplifier tuned to 50 kHz. Optical bandpass filters were used to block the pump waves from reaching the detector.

In a first set of experiments, phase coding of the pump waves was switched off so that BDGs could be formed over the entire length of the fiber under test. Figure 4(a) shows the normalized output lock-in signal $V$ as a function of probe wavelength for a 50-cm-long fiber, with $\Omega$ set to $\Omega_B$. The lock-in signal exhibits multiple sharp peaks at specific probe wavelengths, which represent coupling of the probe wave to cladding modes of consecutive odd orders between 11 and 21. The calculated reflectivity spectrum according to Eqs. (1) and (3) is presented as well. The observed wavelengths of probe wave coupling to the cladding modes agree very well with predictions. Quantitative comparison among the peaks' strengths could not be made due to residual wavelength dependence of the tunable probe laser source, fiber amplifiers, and optical bandpass filters. Figure 4(b) shows a two-dimensional scan of the normalized $V(\Omega, \Delta\omega)$ in the same fiber, near the peak of coupling to cladding mode, $m = 17$. The linewidth of the coupling spectrum with respect to changes in $\Omega$ is $2\pi \times 37$ MHz, in accordance with the Brillouin linewidth. The full width at half-maximum with respect to $\Delta\omega$ is $2\pi \times 260$ MHz, somewhat broader than the predicted width of $2\pi \times 190$ MHz. The difference could be due to inhomogeneity in the local cladding diameter (see also Fig. 5 and Section 3 (Discussion) below).

Next, phase coding of the pump waves was activated to confine the BDGs to short correlation peaks. The period of the phase code was sufficiently long so that only a single correlation peak was formed along the short fiber under test during each acquisition. The normalized signal $V$ was measured as a function of $\Delta\omega$ and correlation peak position $z$, along 2 m of bare fiber, with $\Omega = \Omega_B$. The frequency offset $\Delta\omega$ in each $z$ was scanned in $2\pi \times 200$ MHz increments within a $2\pi \times 7$ GHz-wide range. The integration time of the lock-in signal for each data point $V(z, \Delta\omega)$ was 0.7 s. The signal-to-noise ratio of the lock-in signal measurement at the probe frequencies of maximum coupling was 15 dB. Measurement results are shown in Fig. 5(a). Coupling to cladding mode $m = 17$ is observed in all locations. The offset $\Delta\omega_{opt}^{(17)}(z)$ of maximum coupling changes along the fiber, probably due to local variations in the cladding diameter. Measurements of $\Delta\omega_{opt}^{(17)}$ were repeatable:

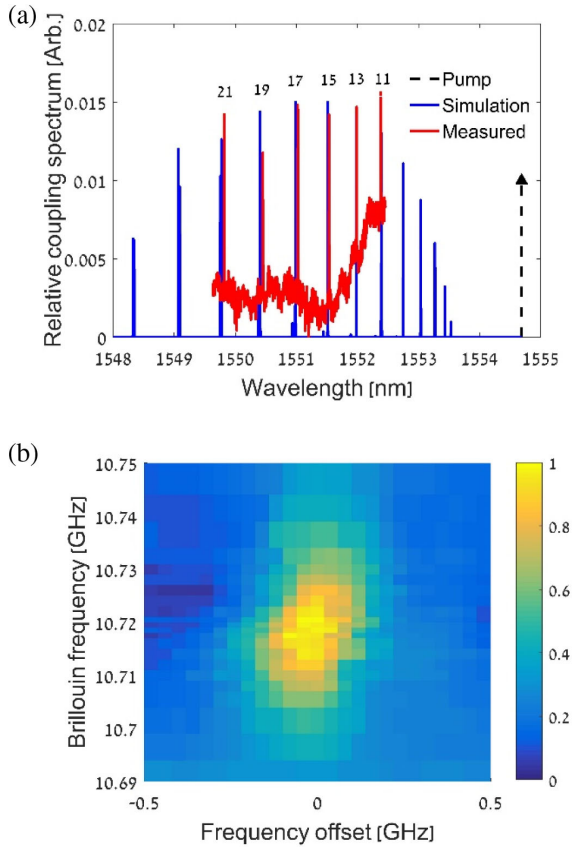

**Fig. 4.** BDG coupling to cladding modes. (a) Red, measured, normalized lock-in modulation signal of the output probe wave, as a function of the wavelength. The pump's wavelength is noted in a black dashed arrow. The frequency difference between the two pump waves was adjusted to match $\Omega_B$. BDGs were formed along an entire 50-cm-long fiber under test. Multiple peaks are observed corresponding to the coupling of the probe wave to cladding modes of odd orders between 11 and 21. Mode orders are noted above the spectral peaks. Blue, theoretical reflectivity spectrum, calculated using Eqs. (1) and (3). The measured wavelengths of peak coupling agree very well with calculations. (b) Measured normalized lock-in signal as a function of the detuning $\Omega/(2\pi)$ between the two pump waves and the difference $\Delta\omega/(2\pi)$ between the pump and probe waves in the vicinity of the coupling peak to cladding mode $m = 17$. The offset $\Delta\omega/(2\pi)$ of maximum coupling is designated as the zero point for convenience.

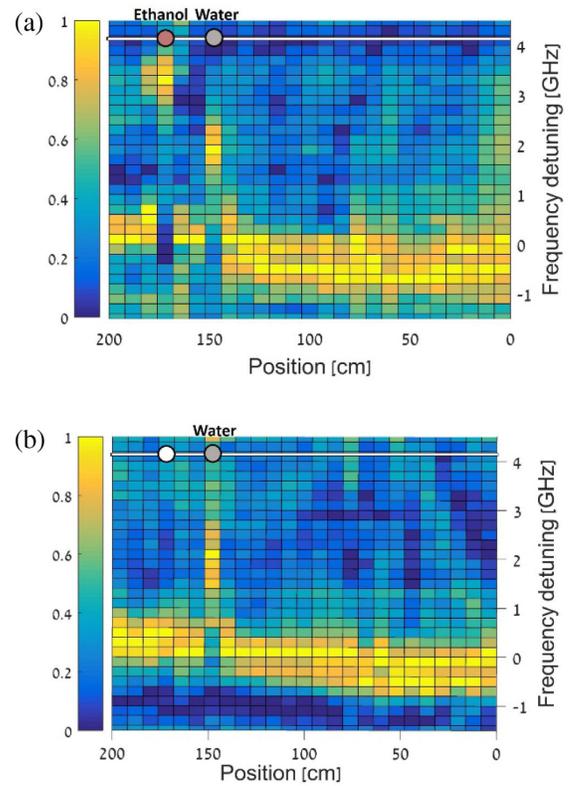

**Fig. 5.** Distributed sensing outside the fiber cladding. (a) Measured, normalized lock-in modulation signal of the output probe wave as a function of the frequency detuning $\Delta\omega/(2\pi)$ between pump and probe and the position $z$ of a localized BDG. Data in each $z$ is normalized to a maximum of unity. Coupling to cladding mode $m = 17$ is observed in all positions. The bare fiber under test is kept in air, except for two 8-cm-wide sections that were immersed in water and ethanol (see legend). The optimal frequency offsets $\Delta\omega^{(17)}_{opt}/(2\pi)$ at the two locations are shifted with respect to the baseline by 1.9 and 3.2 GHz, respectively. (b) Same as panel (a), with the short fiber section at $z = 175$ cm taken out of ethanol (see legend). The optimal frequency offset at that location returned to its value for bare fiber. The second short section at $z = 150$ cm remained in water.

the standard deviation among 50 successive acquisitions at a given location was only $\pm 2\pi \times 75$ MHz.

Two 8-cm-long sections of the fiber under test were immersed in liquids: one in water ($z = 150$ cm) and another in ethanol ($z = 175$ cm). Both regions are clearly identified in the measurements. The local values of $\Delta\omega^{(17)}_{opt}$ are offset by $2\pi \times 1.9$ GHz and $2\pi \times 3.2$ GHz in the two positions, respectively [Fig. 5(a)]. The observed shifts are in agreement with predictions (see Fig. 2) and clearly distinguish between the two liquids. Figure 5(b) shows a repeating measurement of $V(\Delta\omega, z)$ in which only water remained, while the ethanol was allowed to evaporate. The local $\Delta\omega^{(17)}_{opt}$ at $z = 175$ cm returned to its base value for the bare fiber in air, whereas the optimal frequencies elsewhere were unchanged.

As an illustrative control experiment, Fig. 6 shows a standard B-OCDA trace of the coupling of power between the two pump waves as a function of $z$ and $\Omega$, with both liquids in place and the probe wave turned off. The local Brillouin gain spectra in the optical core mode cannot detect the presence of liquids outside the cladding, as anticipated.

## 3. DISCUSSION

Dynamic, random-access coupling of light to cladding modes of a standard single-mode fiber was demonstrated based on BDGs. Optical waves were launched and detected in the core mode only. No permanent gratings were necessary. The principle was employed in a first demonstration of a distributed cladding mode optical fiber sensor: outside media were mapped over 2 m range, with 8 cm resolution. Measurements clearly identified short sections of fiber that were immersed in water and ethanol. The results significantly extend upon previous cladding mode fiber sensors, which have been restricted to point measurements. The experimental setup provides sufficient signal despite the poor spatial overlap between core and cladding modes. In contrast with distributed optomechanical sensing setups [25–32], the proposed method observes the optical properties of outside media rather than their mechanical impedance. The obtained spatial resolution is 2–3 orders of magnitude higher than those of optomechanical sensors [28,29].

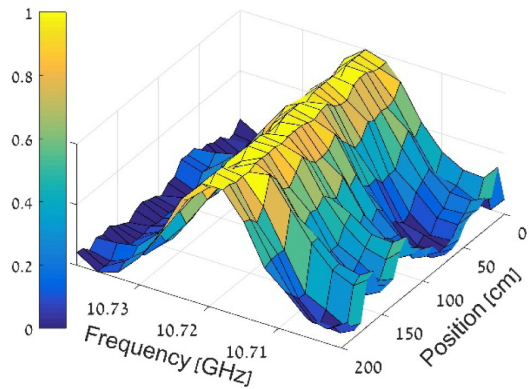

**Fig. 6.** Control experiment. Measured, normalized SBS gain map in the core optical mode of the fiber, as a function of position $z$ and detuning $\Omega/(2\pi)$ between the two pump waves. Two short fiber segments were immersed in liquids as in Fig. 5(a). Standard Brillouin analysis cannot identify the presence of liquids outside the cladding, as anticipated.

---

The sensitivity of coupling spectra to local changes in the refractive index $n_{ext}$ outside the cladding is very weak when $n_{ext}$ is close to unity, but becomes large when the surrounding index approaches that of silica. The experimental error of $\pm 2\pi \times 75$ MHz in the measurements of $\Delta\omega_{opt}^{(m)}$ corresponds to index uncertainty of $\pm 4e$-3 RIU when $n_{ext}$ equals 1.3 RIU. The uncertainty is improved to $\pm 4e$-4 RIU for $n_{ext}$ of 1.4 RIU. Frequency drifts between the free-running laser sources of pumps and probe during the acquisition duration could contribute to the measurement uncertainty. Locking of both lasers to a common frequency comb source can stabilize the frequency difference. The sensing concept is limited to $n_{ext}$ below that of silica. When the surrounding index is higher, the cladding modes are no longer guided.

The effective index of the cladding modes also depends on the cladding diameter. Simulations suggest that $\Delta\omega_{opt}^{(m)}(z)$ changes by 250 MHz per 0.1 µm variation in diameter. The observed deviations in $\Delta\omega_{opt}^{(m)}(z)$ along the fiber under test in Fig. 5(a) are probably due to such local geometric variations, which may be calibrated in advance. Changes in temperature and axial strain affect $n_{core}$ and $n_{clad}^{(m)}$ in a similar manner. In the first order, therefore, temperature and strain should induce little change in $\Delta\omega_{opt}^{(m)}(z)$. Second-order corrections can be calibrated through standard B-OCDA measurements of $\Omega_B(z)$ [58–61].

The measurement range and the number of resolution points are restricted by two considerations: small signal due to the weak BDG coupling between core and cladding modes, and noise due to residual, off-peak Brillouin interactions. The net acquisition duration required to obtain $\Delta\omega_{opt}^{(m)}(z)$ in a single location with 8 cm resolution was 25 s (though the experimental duration was doubled by latencies of laboratory equipment). The measurement range in such conditions may be extended up to a few hundreds of resolution points, or a few tens of meters, before acquisition durations become prohibitively long. The BDG reflectivity scales with its length squared [see Eq. (3)]. Therefore, if the resolution is degraded by some factor $\alpha$, the number of resolution points that can be addressed within the same experimental duration could be increased by $\alpha$ squared. Thousands of resolution points might be addressed with submeter resolution. The signal may also be enhanced by using fibers with smaller cladding. The transverse efficiency is expected to improve twofold in commercially available fibers with 80-µm-diameter cladding.

The measurement signal-to-noise ratio over long fibers would be degraded by the accumulation of residual, off-peak Brillouin interactions [61]. Several mitigation schemes have been developed as part of B-OCDA setups and may be applicable to distributed cladding mode sensing as well. These include time-gating of pump pulses [61,62,64], and dual-layer hierarchal coding of both amplitude and phase [65]. In addition, off-peak Brillouin interactions may be suppressed by changing $\Omega_B(z)$ along the fiber. Specific methods include local heating, position-dependent strain, and concatenation of dissimilar fibers.

The experiments reported in this work were performed using bare fibers. The employment of uncoated fibers outside the research laboratory is unrealistic. However, the proposed sensing concept could be applicable to coated fibers as well, if the refractive index of the coating is lower than that of silica. Fibers with low-index coating are commercially available, such as TrueClad fibers from OFS-Furukawa [66]. A coating layer may further degrade the spatial overlap between the core and cladding modes; therefore, a thin layer would be advantageous. On the other hand, coating also presents opportunities: the index and thickness of the coating layer may be optimized to increase the sensitivity of $\Delta\omega_{opt}^{(m)}(z)$ to changes in $n_{ext}$ within a specific range of interest.

## 4. CONCLUSION

In conclusion, dynamic random-access coupling of light to the cladding modes of a standard fiber has been proposed and demonstrated for the first time. The concept overcomes a decades-long challenge of optical fiber sensors: the high-resolution, distributed optical analysis of media outside the cladding boundary. Future work would address measurements outside coated fibers and increasing the range and number of resolution points. Potential applications include leak detection in critical infrastructure and process monitoring in the petrochemical industry, desalination plants, and food and beverage production.

**Funding.** European Research Council (H2020-ERC-2015-STG 679228 (L-SID)); Israel Science Foundation (1665-14).

**Acknowledgment.** The authors thank Dr. Nachum Gorbatov from Tel-Aviv University for his assistance with the characterization of fibers under test. H. Hagai Diamandi is grateful to the Azrieli Foundation for the award of an Azrieli Fellowship.

See Supplement 1 for supporting content.

# Distributed cladding mode fiber-optic sensor: supplementary material

G. Bashan, Yosef London, H. Hagai Diamandi, Avi Zadok*

*Faculty of Engineering and Institute for Nano-Technology and Advanced Materials, Bar-Ilan University, Ramat-Gan 5290002, Israel *Corresponding author: Avinoam.Zadok@biu.ac.il*

This document provides supplementary information to "Distributed cladding mode fiber-optic sensor," https://doi.org/10.1364/OPTICA.377610. Mathematical analysis of random-access, dynamic and localized coupling of light the cladding modes of a standard optical fiber is provided. Coupling is based on the stimulation of Brillouin dynamic gratings by two pump tones that counter-propagate in the core mode of the fiber. A third optical probe wave may be reflected by the dynamic grating into a counter-propagating cladding mode. Phase matching and spatial overlap considerations are discussed, and expressions for the magnitude and spectrum of coupling are obtained. The localization of dynamic gratings through phase coding of the pump waves is briefly reviewed.

## 1. Coupling to cladding modes of a standard single-mode fiber using Brillouin dynamic gratings

Let $E_{\text{pump1}}(r,z,t)$ denote the optical field of a first continuous pump wave, propagating in the single core mode of a standard fiber in the positive $\hat{\mathbf{z}}$ direction:

$$E_{\text{pump1}}(r,z,t) = A_{\text{pump1}}(z)\exp(jk_{\text{pump1}}z - j\omega_{\text{pump1}}t)u_{\text{core}}(r) + c.c \quad (S1)$$

Here $r$ and $z$ denote the radial and axial coordinates within the fiber, $t$ stands for time, $\omega_{\text{pump1}}, k_{\text{pump1}}$ are the temporal frequency and axial wavenumber of the optical field, and $u_{\text{core}}(r)$ (in units of m$^{-1}$) is the transverse profile of the core mode. The transverse profile is radially-symmetric, and normalized so that $2\pi\int_0^\infty |u_{\text{core}}(r)|^2 r\mathrm{d}r = 1$. Last, $A_{\text{pump1}}(z)$ (in V) represents the local complex magnitude of the first pump wave. A second, co-polarized and continuous optical pump field is counter-propagating in the core mode, in the negative $\hat{\mathbf{z}}$ direction:

$$E_{\text{pump2}}(r,z,t) = A_{\text{pump2}}(z)\exp(-jk_{\text{pump2}}z - j\omega_{\text{pump2}}t)u_{\text{core}}(r) + c.c \quad (S2)$$

In Supplementary Equation (S2), $A_{\text{pump2}}(z)$ represents the complex magnitude of the second pump wave, and $k_{\text{pump2}}$ is its wavenumber. The optical frequency of the second pump wave is given by $\omega_{\text{pump2}} = \omega_{\text{pump1}} - \Omega$, where $\Omega$ is close to the Brillouin frequency shift $\Omega_B$ in the fiber. We denote the effective index of the core mode as $n_{\text{core}}$, so that $k_{\text{pump1,2}} = n_{\text{core}}\omega_{\text{pump1,2}}/c$ where $c$ is the speed of light in vacuum. We assume that $n_{\text{core}}$ is the same for all frequencies of interest.

Backward stimulated Brillouin scattering (SBS) interaction between the two pump fields generates a longitudinal acoustic wave of density fluctuations $\Delta\rho(r,z,t)$, which is co-propagating with $E_{\text{pump1}}$ [S1]:

$$\Delta\rho(r,z,t) = B(z,\Omega)\exp(jqz - j\Omega t)u_{\text{ac}}(r) + c.c. \quad (S2)$$

Here $q = k_{\text{pump1}} + k_{\text{pump2}}$ is the acoustic wavenumber and $u_{\text{ac}}(r)$ is the transverse profile of a longitudinal acoustic mode that is guided by the core of the fiber, normalized to $2\pi\int_0^\infty |u_{\text{ac}}(r)|^2 r\mathrm{d}r = 1$. The transverse profile of the electrostrictive force induced by the two pump waves is radially-symmetric, hence the stimulated acoustic mode must maintain the same symmetry. The magnitude of the acoustic wave, in units of kg×m$^{-2}$, is given by [S1]:

$$B(z,\Omega) = \varepsilon_0\gamma_e q^2 Q_{\text{core}} \frac{1}{\Omega_B^2 - \Omega^2 - j\Omega\Gamma_B} A_{\text{pump1}}(z)A_{\text{pump2}}^*(z) \quad (S3)$$

In Supplementary Equation (S4) $\varepsilon_0$ is the vacuum permittivity, $\gamma_e$ is the electrostrictive constant of silica, and $\Gamma_B \approx 2\pi \times 30$ MHz represents the Brillouin linewidth in silica. $Q_{core}$ [m$^{-1}$] denotes the spatial overlap integral between the transverse profile of the optical intensity in the core mode and that of the acoustic mode:

$$Q_{core} \equiv 2\pi \int_0^\infty |u_{core}(r)|^2 u_{ac}^*(r) r\,dr \quad (S4)$$

Consider next a third continuous optical probe wave, which is co-propagating with $E_{pump1}$ in the core optical mode of the fiber:

$$E_{probe}(r,z,t) = A_{probe}(z)\exp(jk_{probe}z - j\omega_{probe}t)u_{core}(r) + c.c \quad (S5)$$

The complex magnitude of the probe wave is denoted by $A_{probe}(z)$, its optical frequency is $\omega_{probe} > \omega_{pump1}$, and its wavenumber $k_{probe}$ equals $k_{probe} = n_{core}\omega_{probe}/c$. The combination of the probe optical field and the stimulated acoustic wave is associated with a nonlinear polarization term at optical frequency $\omega_{probe} - \Omega$, due to photo-elasticity [S1]:

$$P_{NL}(r,z,t) = \varepsilon_0 \frac{\gamma_e}{\rho_0} B^*(z) A_{probe}(z) \times$$
$$\times u_{ac}^*(r) u_{core}(r) \exp[j(k_{probe}-q)z - j(\omega_{probe}-\Omega)t] + c.c. \quad (S6)$$

Here $\rho_0$ is the density of silica. Note that $(k_{probe} - q) < 0$, hence the nonlinear polarization represents a wave perturbation propagating in the negative $\hat{z}$ direction. The above nonlinear polarization can lead to scattering of the probe wave into a counter-propagating, fourth optical field of frequency $\omega_{probe} - \Omega$, provided that its wavenumber matches that of Supplementary Equation (S7). This requirement can be satisfied in the $m^{th}$-order cladding mode of the fiber, with effective index $n_{clad}^{(m)} < n_{core}$, if the following condition is met:

$$k_{clad}^{(m)} = \frac{n_{clad}^{(m)}}{c}(\omega_{probe}-\Omega) = q - k_{probe} =$$
$$\frac{n_{core}}{c}(\omega_{pump1} + \omega_{pump1} - \Omega - \omega_{probe}) \quad (S8)$$

Based on Supplementary Equation (S8), Brillouin dynamic grating (BDG) coupling to the cladding mode is optimal for the following optical frequency of the probe wave:

$$\omega_{probe,opt}^{(m)} = \frac{2n_{core}\omega_{pump1} - (n_{core} - n_{clad}^{(m)})\Omega}{n_{core} + n_{clad}^{(m)}} \approx \frac{2n_{core}}{n_{core} + n_{clad}^{(m)}}\omega_{pump1} \quad (S7)$$

which is also Equation (1) in the Main Text. Note that the second term in the numerator of Supplementary Equation (S9) is smaller than the first by seven orders of magnitude, hence the approximation made is a very good one. The frequency detuning between the co-propagating pump and probe waves at optimal coupling is given by:

$$\Delta\omega_{opt}^{(m)} \equiv \omega_{probe,opt}^{(m)} - \omega_{pump1} = \frac{n_{core} - n_{clad}^{(m)}}{n_{core} + n_{clad}^{(m)}}\omega_{pump1} \quad (S10)$$

Supplementary Fig. 1 shows the calculated $n_{clad}^{(m)}$ for a 125 μm-diameter fiber in air [S2], and the optimal wavelength offset between the probe and a pump waves at 1550 nm, as functions of the cladding mode order $m$.

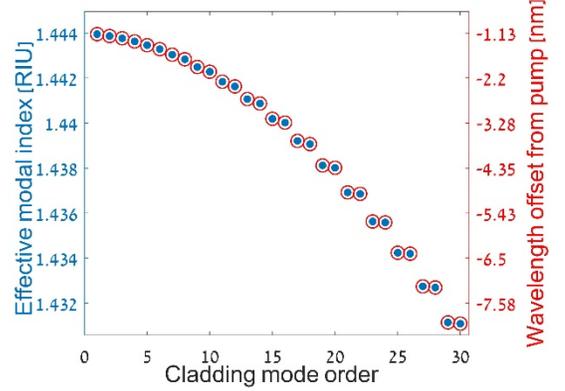

Supplementary Fig. 1. – Calculated effective index $n_{clad}^{(m)}$ (left axis), and wavelength offset between pumps and probe at maximum coupling of the probe wave to cladding modes (right axis, based on Supplementary Equation 10), as a function of cladding mode order $m$. A bare fiber with air outside the cladding is assumed. The cladding diameter is 125 μm.

The scattered field in the cladding mode may be expressed as:
$$E_{clad}^{(m)}(r,z,t) =$$
$$A_{clad}^{(m)}(z)\exp[-jk_{clad}^{(m)}z - j(\omega_{probe}-\Omega)t]u_{clad}^{(m)}(r) + c.c \quad (S11)$$

where $A_{clad}^{(m)}(z)$ is the complex magnitude of the scattered field and $u_{clad}^{(m)}(r)$ is the normalized transverse profile of the $m^{th}$-order cladding mode. The transverse profile of the cladding mode follows the radial symmetry of the optical core mode and the acoustic mode. The coupled nonlinear wave equations for the evolution of $A_{probe}(z)$ and $A_{clad}^{(m)}(z)$ take up the following form:

$$\frac{dA_{probe}(z)}{dz} = j\frac{\varepsilon_0\gamma_e^2 q^2 \omega_{probe} Q_{core} Q_{clad/core}^{(m)}}{2nc\rho_0} \times$$
$$\frac{A_{pump1}(z)A_{pump2}^*(z)}{\Omega_B^2 - \Omega^2 - j\Omega\Gamma_B} A_{clad}^{(m)}(z)\exp(-j\Delta k_{BDG}^{(m)}z) \quad (S12)$$

$$\frac{dA_{clad}^{(m)}(z)}{dz} = -j\frac{\varepsilon_0\gamma_e^2 q^2 \omega_{probe} \left(Q_{core} Q_{clad/core}^{(m)}\right)^*}{2nc\rho_0} \times$$
$$\frac{A_{pump1}^*(z)A_{pump2}(z)}{\Omega_B^2 - \Omega^2 + j\Omega\Gamma_B} A_{probe}(z)\exp(j\Delta k_{BDG}^{(m)}z) \quad (S13)$$

Here we use $n_{clad}^{(m)} \approx n_{core} \equiv n$ and $(\omega_{probe} - \Omega) \approx \omega_{probe}$. The term $Q_{clad/core}^{(m)}$ [m$^{-1}$] stands for the spatial overlap integral between the transverse profiles of the core, cladding and acoustic modes:

$$Q_{clad/core}^{(m)} \equiv 2\pi \int_0^\infty u_{ac}(r) u_{clad}^{(m)}(r) u_{core}^*(r) r\,dr \quad (S14)$$

Lastly, the wavenumber mismatch term in Supplementary Equations (S12) and (S13) is defined as:

$$\Delta k_{\text{BDG}}^{(m)} \equiv q - k_{\text{probe}} - k_{\text{clad}}^{(m)} = \frac{n_{core} + n_{\text{clad}}^{(m)}}{c}\left(\omega_{\text{probe}} - \omega_{\text{probe,opt}}^{(m)}\right) \quad \text{(S15)}$$

For brevity, we rearrange the coefficients of Supplementary Equations (S12) and (S13):

$$-\text{j}\frac{\varepsilon_0 \gamma_e^2 q^2 \omega_{\text{probe}} Q_{\text{core}} Q_{\text{clad/core}}^{(m)}}{2nc\rho_0}\frac{1}{\Omega_B^2 - \Omega^2 - \text{j}\Omega\Gamma_B} \approx$$

$$-\text{j}\frac{\varepsilon_0 \gamma_e^2 q^2 \omega_{\text{probe}} Q_{\text{core}} Q_{\text{clad/core}}^{(m)}}{2nc\rho_0}\frac{1}{-\text{j}\Omega_B \Gamma_B}\frac{1}{1+\text{j}2\Delta\Omega/\Gamma_B}$$

$$\frac{\varepsilon_0 \gamma_e^2 q^2 \omega_{\text{probe}} Q_{\text{core}} Q_{\text{clad/core}}^{(m)}}{2nc\rho_0 \Omega_B \Gamma_B}\frac{1-\text{j}2\Delta\Omega/\Gamma_B}{1+(2\Delta\Omega/\Gamma_B)^2} = \quad \text{(S16)}$$

$$C_{\text{BDG},0} Q_{\text{core}} Q_{\text{clad/core}}^{(m)} \frac{1-\text{j}2\Delta\Omega/\Gamma_B}{1+(2\Delta\Omega/\Gamma_B)^2}$$

where $\Delta\Omega \equiv \Omega_B - \Omega$, and:

$$C_{\text{BDG},0} \equiv \frac{\varepsilon_0 \gamma_e^2 q^2 \omega_{\text{probe}}}{2nc\rho_0 \Omega_B \Gamma_B} \quad \text{(S17)}$$

We assume further that the two pump waves are undepleted in the short fiber under test. With the above definitions, we may rewrite Supplementary Equations (S12) and (S13):

$$\frac{\text{d}A_{\text{probe}}(z)}{\text{d}z} = -C_{\text{BDG},0} Q_{\text{core}} Q_{\text{clad/core}}^{(m)} A_{\text{pump1}} A_{\text{pump2}}^* \times$$

$$\frac{1-\text{j}2\Delta\Omega/\Gamma_B}{1+(2\Delta\Omega/\Gamma_B)^2} A_{\text{clad}}^{(m)}(z) \exp\left(-\text{j}\Delta k_{\text{BDG}}^{(m)} z\right) \quad \text{(S18)}$$

$$\frac{\text{d}A_{\text{clad}}^{(m)}(z)}{\text{d}z} = -C_{\text{BDG},0}\left(Q_{\text{core}} Q_{\text{clad/core}}^{(m)}\right)^* A_{\text{pump1}}^* A_{\text{pump2}} \times$$

$$\frac{1+\text{j}2\Delta\Omega/\Gamma_B}{1+(2\Delta\Omega/\Gamma_B)^2} A_{\text{probe}}(z) \exp\left(\text{j}\Delta k_{\text{BDG}}^{(m)} z\right) \quad \text{(S19)}$$

Due to the short lengths of fiber used, and the small spatial overlap between core and cladding modes $Q_{\text{clad/core}}^{(m)}$, the coupling of power between $E_{\text{probe}}$ and $E_{\text{clad}}^{(m)}$ is very weak. Although $|A_{\text{probe}}|^2 \ll |A_{\text{pump1,2}}|^2$, we may still assume that changes in $A_{\text{probe}}$ remain small (although nonzero). At that limit, the magnitude of the optical field that is coupled into the cladding mode by a BDG of length $L_{\text{BDG}}$ is approximately given by:

$$A_{\text{clad,out}}^{(m)} \approx C_{\text{BDG},0}\left(Q_{\text{core}} Q_{\text{clad/core}}^{(m)}\right)^* A_{\text{pump1}}^* A_{\text{pump2}}$$

$$\times \frac{1+\text{j}2\Delta\Omega/\Gamma_B}{1+(2\Delta\Omega/\Gamma_B)^2} A_{\text{probe}}$$

$$\times \exp\left(\text{j}\frac{\Delta k_{\text{BDG}}^{(m)} L_{\text{BDG}}}{2}\right) L_{\text{BDG}} \quad \text{(S20)}$$

$$\times \text{sinc}\left(\frac{\Delta k_{\text{BDG}}^{(m)} L_{\text{BDG}}}{2}\right)$$

The optical power coupled to the cladding mode is given by $P_{\text{clad,out}}^{(m)} = 2nc\varepsilon_0 \left|A_{\text{clad,out}}^{(m)}\right|^2$ (see [S3]):

$$P_{\text{clad,out}}^{(m)} \approx \frac{2\varepsilon_0 nc C_{\text{BDG},0}^2 |Q_{\text{core}}|^2 \left|Q_{\text{clad/core}}^{(m)}\right|^2 |A_{\text{pump1}}|^2 |A_{\text{pump2}}|^2}{1+(2\Delta\Omega/\Gamma_B)^2}$$

$$\times |A_{\text{probe}}|^2 L_{\text{BDG}}^2 \text{sinc}^2\left(\frac{\Delta k_{\text{BDG}}^{(m)} L_{\text{BDG}}}{2}\right)$$

$$= \frac{C_{\text{BDG},0}^2 |Q_{\text{core}}|^2 \left|Q_{\text{clad/core}}^{(m)}\right|^2 |A_{\text{pump1}}|^2 |A_{\text{pump2}}|^2}{1+(2\Delta\Omega/\Gamma_B)^2}$$

$$\times \text{sinc}^2\left(\frac{\Delta k_{\text{BDG}}^{(m)} L_{\text{BDG}}}{2}\right) L_{\text{BDG}}^2 P_{\text{probe}} \quad \text{(S21)}$$

$$= \left(\frac{C_{\text{BDG},0}}{2nc\varepsilon_0}\right)^2 \frac{|Q_{\text{core}}|^2 \left|Q_{\text{clad/core}}^{(m)}\right|^2 P_{\text{pump1}} P_{\text{pump2}}}{1+(2\Delta\Omega/\Gamma_B)^2}$$

$$\times \text{sinc}^2\left(\frac{\Delta k_{\text{BDG}}^{(m)} L_{\text{BDG}}}{2}\right) L_{\text{BDG}}^2 P_{\text{probe}}$$

Here $P_{\text{probe}} = 2nc\varepsilon_0 |A_{\text{probe}}|^2$ is the input power of the probe wave, and $P_{\text{pump1,2}} = 2nc\varepsilon_0 |A_{\text{pump1,2}}|^2$ are the input powers of the two pump waves. The reflected power obtains a maximum value when the frequency offset $\Omega$ between the two pump waves matches exactly the Brillouin frequency shift $\Omega_B$ ($\Delta\Omega = 0$), and the frequency of the probe equals $\omega_{\text{probe,opt}}^{(m)}$ ($\Delta k_{\text{BDG}}^{(m)} = 0$). The reflectivity bandwidth with respect to $\Omega$ equals the Brillouin linewidth $\Gamma_B$. The bandwidth with respect of $\omega_{\text{probe}}$ is inversely proportional to the BDG length $L_{\text{BDG}}$. The maximum BDG reflectivity into the cladding mode may be expressed as:

$$R_{\text{max}}^{(m)} \equiv \frac{P_{\text{clad,out}}^{(m)}}{P_{\text{probe}}}\bigg|_{\Delta\Omega = \Delta k_{\text{BDG}}^{(m)} = 0} \approx$$

$$\left(\frac{C_{\text{BDG},0}}{2nc\varepsilon_0}\right)^2 |Q_{\text{core}}|^2 \left|Q_{\text{clad/core}}^{(m)}\right|^2 P_{\text{pump1}} P_{\text{pump2}} L_{\text{BDG}}^2 \quad \text{(S22)}$$

$$= D_{\text{BDG},0}^2 |Q_{\text{core}}|^2 \left|Q_{\text{clad/core}}^{(m)}\right|^2 P_{\text{pump1}} P_{\text{pump2}} L_{\text{BDG}}^2,$$

where the coefficient $D_{\text{BDG},0}$ (in units of m×W$^{-1}$) is defined as:

$$D_{\text{BDG},0} \equiv \frac{\gamma_e^2 q^2 \omega_{\text{probe}}}{4n^2 c^2 \rho_0 \Omega_B \Gamma_B} \approx \frac{\gamma_e^2}{2c^3 nv\rho_0 \Gamma_B} \omega^2 \quad \text{(S23)}$$

Here we used $q = \Omega/v$ with $v$ the velocity of longitudinal acoustic waves, and we approximated $\Omega \approx \Omega_B \approx 2n\omega_{pump1} v/c$ and $\omega_{pump1} \approx \omega_{probe}$ denoted by $\omega$.

The power reflectivity to the cladding mode is weaker than that of a BDG in a polarization maintaining fiber by a transverse efficiency factor $\eta^{(m)} \equiv \left|Q^{(m)}_{clad/core}\right|^2 / \left|Q_{core}\right|^2$. Supplementary Fig. 2 shows the numerically calculated $\eta^{(m)}$ as a function of cladding mode order $m$. The cladding radius was taken as 125 μm, the core radius was 4.1 μm, and the mode-field diameter of the core mode $|u_{core}(r)|^2$ was 9.7 μm. Calculations were repeated twice: first under the assumption that the transverse profile of the acoustic mode $u_{ac}(r)$ is proportional to that of optical intensity in the core mode $|u_{core}(r)|^2$, and again for uniform $u_{ac}(r)$ within the core. The former model follows the transverse profile of the electrostrictive driving force, whereas the latter matches that of permanent fiber Bragg gratings [S2]. Differences between the two sets of results are small.

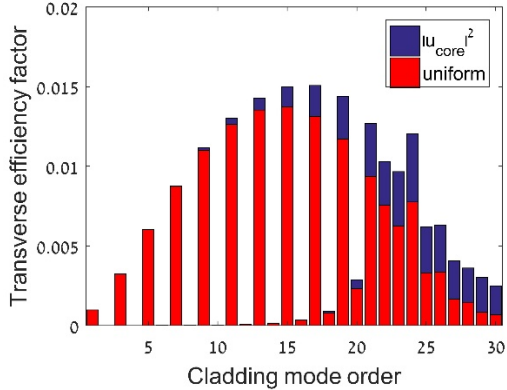

Supplementary Fig. 2. – Calculated transverse efficiency factor of Brillouin dynamic grating coupling as a function of cladding mode order. Blue: the transverse profile of the acoustic mode was assumed to be proportional to the optical intensity profile of the optical core mode: $u_{ac}(r) \propto |u_{core}(r)|^2$. Red: the transverse profile of the acoustic mode $u_{ac}(r)$ was assumed to be uniform across the core.

The analysis suggests that BDG coupling is the most efficient for odd cladding mode orders between 13 and 19 [S2]. Note that the even cladding modes are characterized by zero optical field on the fiber axis. The radial profiles of the cladding mode fields oscillate with periods that become shorter as the modal order increases. For low-order even modes, the field remains very weak throughout the extent of the core and the transverse efficiency of Brillouin dynamic grating coupling vanishes accordingly. For higher-order even modes, radial variations in the field profile are gradually pushed into the core, giving rise to somewhat larger transverse overlap with the dynamic gratings.

The highest transverse efficiency is expected for mode $m$ = 17. However, even $\eta^{(17)}$ is only about 1.5%. The reflectivity $R^{(17)}_{max}$ of few-cm long BDGs into the cladding is weak: on the order of 100 ppm for few Watts of pumps power. Nevertheless, coupling spectra are successfully used in distributed sensing outside the cladding (see Main Text). Due to the difficulty of collecting light from the cladding modes, we monitor the process instead by measuring the changes in the transmitted probe power: $\Delta P_{probe} = -P^{(m)}_{clad,out}$.

## 2. Localization of steady state stimulated Brillouin scattering interactions through phase coding of pump waves

The localization of stimulated Brillouin scattering interactions through the phase coding of two optical waves has been described in detail in several works [S4][S5]. The principle is briefly repeated here for completeness. The reader is referred to a recent review [S6]. In this technique, the two pump fields are no longer continuous. The complex magnitude of pump field $E_{pump1}$ is modulated at its point of entry into the fiber ($z = 0$), so that:

$$A_{pump1}(z,t) = A^{(0)}_{pump1} f\left(t - \frac{z}{v_g}\right) \quad (S24)$$

In Supplementary Equation (S24), $A^{(0)}_{pump1}$ is a constant magnitude, $v_g$ is the group velocity of light in the fiber, and $f(t)$ is a modulation function of the optical source with unity norm: $|f(t)|^2 = 1$. The counter-propagating $E_{pump2}$ is modulated by the same function $f(t)$ at its launch point in the opposite end of the fiber $z = L$, so that:

$$A_{pump2}(z,t) = A^{(0)}_{pump2} f\left(t - \frac{L-z}{v_g}\right) \quad (S25)$$

Here $A^{(0)}_{pump2}$ represents a second constant magnitude. Due to the modulation of the pump waves, the magnitude of the stimulated acoustic density perturbation generally does not reach a steady state. The instantaneous acoustic magnitude is given by [S6]:

$$B(z,t) = j\varepsilon_0 \gamma_e q^2 Q_{core} A^{(0)}_{pump1}\left(A^{(0)}_{pump2}\right)^* \times$$
$$\int_{-\infty}^{t} \exp\left[\Gamma_{ac}(\Omega)\cdot(t-t')\right] \times \quad (S26)$$
$$f\left(t' - \frac{z}{v_g}\right) f\left[t' - \frac{z}{v_g} + \theta(z)\right] dt'$$

In Supplementary Equation (S26) we defined a complex linewidth $\Gamma_{ac}(\Omega) \equiv j(\Omega_B^2 - \Omega^2 - j\Omega\Gamma_B)/2\Omega s$ and a position-dependent time lag $\theta(z) \equiv (2z - L)/v_g$. The complex linewidth reduces to $\frac{1}{2}\Gamma_B$ when $\Omega = \Omega_B$.

Let us denote the auto-correlation function of the modulation waveform $f(t)$ as $C_f(\xi)$, where $\xi$ is a delay variable. Supplementary Equation (S26) suggests that the expectation value of the stimulated acoustic wave at $z$ is closely related with $C_f[\theta(z)]$. Note, however, that the acoustic wave magnitude does not perfectly follow the auto-correlation function, due to the exponential weighing window $\exp\left[\Gamma_{ac}(\Omega)\cdot(t-t')\right]$ that is associated with electrostrictive stimulation.

Following on earlier works [S4-S6], the modulation function $f(t)$ is chosen as a repeating binary phase sequence, with symbol duration $T$ and a period of $N$ bits:

$$f(t) = \sum_n a_n \text{rect}\left(\frac{t-nT}{T}\right) \quad \text{(S27)}$$

Here $\text{rect}(\xi) = 1$ if $|\xi| \leq 0.5$ and equals zero elsewhere, and $a_n$ is the value of bit $n$ in the sequence. The bit duration $T$ is taken to be much shorter than the Brillouin lifetime: $T \ll 1/\Gamma_B$. The values of $a_n$ are those a prefect Golomb code: a class of binary phase sequences that are designed for zero side-lobes of their cyclic auto-correlation functions [S6][S7]. Due to the phase modulation, a correlation peak forms at the center of the fiber $z = L/2$ ($\theta = 0$), where the magnitude of the acoustic wave reaches its steady-state value of Supplementary Equation (S4). The width of the resulting BDG $L_{BDG}$ equals $\frac{1}{2}v_g T$. Periodic, higher-order peaks appear at positions $z = L/2 + M \cdot N \cdot L_{BDG}$, where $M$ is a positive or negative integer.

Outside the correlation peaks, the magnitudes of the stimulated acoustic waves are rapidly fluctuating, as the arguments within the integral of Supplementary Equation (S26) may assume positive or negative values. The expectation values of the acoustic wave magnitudes outside the peak equal zero for all times. Therefore, in principle, measurements of $E_{probe}$ at the end of the fiber may retrieve the local BDG spectrum at the position of the correlation peak only. However, even though off-peak Brillouin interactions are zero on average, their instantaneous magnitudes are non-zero with a finite variance [S6]. Off-peak Brillouin interactions contribute noise to the measurements of BDG coupling spectra [S6].

Measurements of BDGs with phase-coded pump waves are unambiguous for fiber lengths $L$ that are shorter than $N \cdot L_{BDG}$. Since the repeating phase-modulation sequence can be chosen at any length, the range of unambiguous measurements may be arbitrarily long, with no effect on spatial resolution. In many realizations of the concept, the fiber paths leading the two pump waves into the measurement section of interest are deliberately imbalanced, so that a high-order correlation peak ($M \gg 1$) is in overlap with the region of interest [S8]. With this choice, the position of the correlation peak can be conveniently scanned through small-scale variations in the bit duration $T$ [S8].

A distributed map of BDG coupling spectra as a function of position can be obtained by scanning the location of the correlation peak position, one resolution point at a time, and then scanning the optical frequency of the probe wave $\omega_{probe}$ at each position. The technique has been widely employed in distributed Brillouin optical correlation domain analysis (B-OCDA) sensing of temperature and axial strain [S5][S6]. It was also used in BDGs over polarization maintaining fibers, towards sensing, all-optical variable delay lines and microwave-photonic filters [S4][S9-S11]. In this work, phase coding of the two pump waves is used to obtain distributed mapping of local coupling spectra between the optical probe field and cladding modes of the fiber under test (see Main Text).